%% file: lot2.tex
\documentclass[global,twocolumn,final,numbook]{svjour}
\usepackage{graphicx}
\usepackage[pdftex, colorlinks=true, pdfstartview=FitH,linkcolor=blue, citecolor=blue, urlcolor=blue]{hyperref}
\usepackage{color}
\usepackage[ruled,section]{algorithm}
\usepackage{algorithmic}
\usepackage{latexsym}
\usepackage{amssymb,amsfonts,amsbsy,amsmath,amscd}
\usepackage{float}
\usepackage[FIGBOTCAP,normal,bf,tight]{subfigure}
\usepackage{stmaryrd}
\usepackage{enumerate}
\usepackage{upgreek}
\usepackage{xcolor}
\newcommand{\VTK}[1]{\textsf{VTK}#1}

\graphicspath{{./pdf/}{./png/}{./plots/}}
\begin{document}
\title{Two New Contributions to the Visualization of AMR Grids:\\
       I. Interactive Rendering of Extreme-Scale $2$-Dimensional Grids\\
       II. Novel Selection Filters in Arbitrary Dimension}
\author{Gu\'enol\'e \textsc{Harel}\inst{1}
	\and
        Jacques-Bernard \textsc{Lekien}\inst{1}
	\and
	Philippe \textsc{P\'eba\"y}\inst{2}
}
\institute{
        CEA, DAM, DIF, F-91297 Arpajon, France\\
        \email{\{guenole.harel,jacques-bernard.lekien\}@cea.fr}
	\and
	Positiveyes, 84330 Le Barroux, France\\
	\email{philippe.pebay@positiveyes.fr}
\date{} 
}
\def\makeheadbox{} 
\maketitle
\begin{abstract}
We present here the result of continuation work, performed to further
fulfill the vision we outlined in~\cite{harel:17} for the
visualization and analysis of tree-based adaptive mesh refinement
(AMR) simulations, using the \emph{hypertree grid} paradigm which
we proposed.

The first filter presented hereafter implements an adaptive approach in
order to accelerate the rendering of $2$-dimensional AMR grids,
hereby solving the problem posed by the loss of interactivity
that occurs when dealing with large and/or deeply refined meshes.
Specifically, view parameters are taken into account, in
order to: on one hand, avoid creating surface elements that are
outside of  the view area; on the other hand, utilize level-of-detail
properties to cull those cells that are deemed too small to be visible
with respect to the given view parameters.
This adaptive approach often results in a massive increase in
rendering performance.

In addition, two new selection filters provide data analysis
capabilities, by means of allowing for the extraction of those cells
within a hypertree grid that are deemed relevant in some sense, either
geometrically or topologically.
After a description of these new algorithms, we illustrate their use
within the Visualization Toolkit (VTK) in which we implemented them.
This note ends with some suggestions for subsequent work.
\end{abstract}
\keywords{scientific visualization, interactive visualization,
meshing, AMR, mesh refinement, tree-based, octree, VTK, parallel
visualization, large scale visualization, HPC, data analysis}
\tableofcontents
\input{00}

\input{10}
\input{80}
\input{90}
\subsection*{Acknowledgments}
This work was supported by the French Alternative Energies and Atomic
Energy Commission (CEA), Directorate of Military Applications
(DAM). 
We thank J.-C. Frament (Positiveyes) for the helpful discussions in
the context of this work.
\bibliographystyle{plain}
\bibliography{lot2}
\end{document}

%% file: 00.tex
\section{Introduction}
\label{s:introduction}
This short article is a sequel to~\cite{harel:17}, where we 
presented the first systematic treatment of the problems posed by the
visualization and analysis of large-scale, parallel, tree-based
adaptive mesh refinement (AMR) simulations on an Eulerian grid.
In that previous article, we proposed a novel data object for the
Visualization Toolkit (\VTK)~\cite{avila:10}, able to
support all conceivable types of rectilinear, tree-based AMR data sets
not only produced by today's simulation software, but also by what is
foreseeable of tomorrow's extreme-scale simulations.

\begin{figure}[h!]
\centering
\includegraphics[width=0.9\columnwidth]{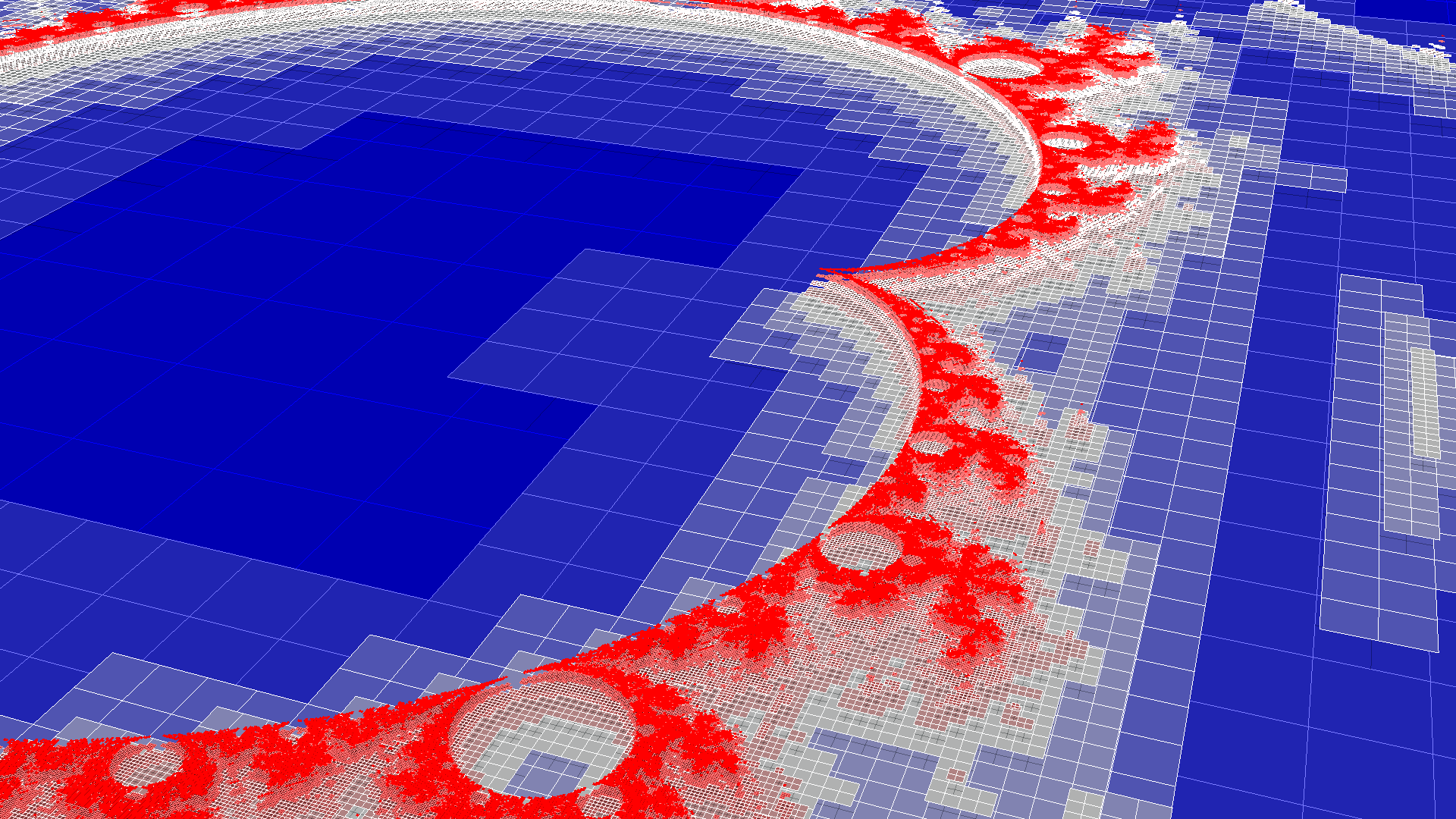}
\caption{Visualization of the different levels of a $2$-dimensional
AMR grid, whose depth levels are raised by increasing elevations in
order to highlight the hierarchic nature of the object.}
\label{fig:AMR_Explose2}
\end{figure}
The concrete result of this earlier work is a novel version
of the \texttt{vtkHyperTreeGrid} data object, differing in many key
aspects from the initial incarnation which we proposed
in~\cite{carrard:imr21}.
The novel version also includes a dozen of visualizations algorithms
(\emph{filters}) operating natively upon the new data object.
An example visualization obtained with this new object is shown in
Figure~\ref{fig:AMR_Explose2}.
Note that this set of filters includes, in particular, a conversion
algorithm to transform a tree-based AMR grid into a fully explicit,
unstructured mesh that can, albeit extremely 
inefficiently, be processed by other visualization techniques not
currently included in the native, hypertree grid filter.
One key aspect of this improved object is that it relies on a
hierarchy of templated traversal objects, called \emph{cursors} and
\emph{supercursors}, allowing for the easy creation of new filters
while retaining the intrinsic performance of our method, in terms
of both execution speed and memory footprint.

Meanwhile, we acknowledged that $2$-dimensional AMR visualization
could be especially challenging, as a result of its requirement that
all leaf cells be rendered.
This constraint makes the interactivity of the visualization process
decrease as input data object size increases.
It is important to note that this problem is further compounded by the
enhanced efficiency, in terms of memory footprint, of our hypertree
grid model.
This elicits a new situation where rendering has become
the bottleneck for the target platforms.

As a result, our next stated goal was to optimize rendering speed, in
order to maintain interactivity when visualizing the largest
possible tree grids that can be contained in memory.
We thus set out to address this urgent need, for which the lack of
existing solution was hindering the AMR visualization and analysis
workflow.
The results of this work are thus presented hereafter.

In addition, we also explain here how we expanded the set of hypertree
grid filters with two novel data selection and extraction filters: one
based on location (i.e., a geometric property), the other on element
Id (i.e., a topological property).
This further supports, in particular, our claim that the new hypertree
grid framework offers enough flexibility and convenience that
algorithms not initially envisioned can be readily added and deployed.

%% file: 10.tex
\section{Method}
\label{s:method}
\subsection{Adaptive Dataset Surface Filter}
\label{s:adaptive-surface-filter}
The purpose of this filter is to extract the outer surface of an
hyper tree grid. 
It is worth noting that the \texttt{Geometry} filter already provides
this capability,~cf.\cite{harel:17}.
However, an additional constraint is added, namely an emphasis on
rendering speed in dimension~$2$.
Our method to address this need is to exploit level-of-detail
properties, culling those parts of the data set that are not
\emph{visible}, i.e., not contained in the visualization frustum
formed by the camera and object positions.
The algorithm explores each tree in the grid, searching for those
faces of the different leaf cells that belong to the surface.
At the end of the computation, the result is a polygonal mesh
(concretely, a \texttt{vtkPolyData} instance) containing an array of
rectangular cells, that can be used to render the surface of the hyper
tree.

The main difference between this adaptive geometry filter and the
non-adaptive one is that it allows for the use of the \emph{renderer}
parameters to reduce the computational work.
Note that currently, this feature is only available when working with
a $2$-dimensional mesh and a parallel projection is set.
When working under these circumstances, the filter computes only the
part of the surface of the grid that is going to be rendered.
This also means that whenever any of the parameters of the
camera associated to the renderer are modified, the output surface
must be subsequently recomputed.

In order to traverse a $2$-dimensional hypertree grid, the algorithm
uses a \emph{geometric cursor}.
We briefly recall here that this cursor provides information about the
size and the coordinates of the cells of the tree and the number of
children, cf.\cite{harel:17} for details.
The algorithm then recursively processes the entire tree, cell by cell
starting from the root level and using a depth-first search (DFS)
order.
In order to cull those cells that are below a certain size in the
rendering window, the DFS traversal is not allowed to descend below
the following maximum depth:
\begin{equation}
\label{eq:level-max}
\delta_{\max}(w,z,s,f) = \frac{\log(wz) - \log{s}}{\log{f}}
\end{equation}
where $w$, $z$, $s$ and $f$ respectively denote the size,
zoom-in factor, scale of the rendering window, and the branching
factor of the hypertree grid.

If the cell to be processed is outside of the rendering
area, it is discarded, which results in reducing the amount of
computational work to be performed by the algorithm. 
Non-leaf cells only need to recursively process their children.
Leaf cells are processed, unless they are masked, in which case they
shall not participate in the output surface.
If a cell is not masked, its geometry and topology are computed, and
these are added to the list of points and cells defining the polygonal
mesh output.

When the input is a $3$-dimensional hyper tree grid we find a different
situation, as we only want to generate the outer surface we need to
omit internal surfaces: in this case, the same method as that used by
the \texttt{Geometry} filter is used.

\begin{figure}[ht!]
\centering
\begin{minipage}[t]{0.9\columnwidth}
\includegraphics[width=\columnwidth]{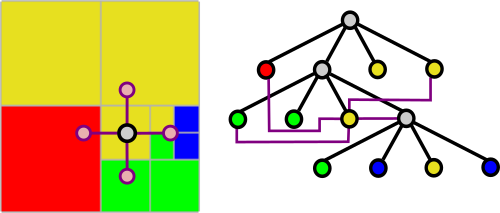}
\end{minipage}
\caption{Von Neumann supercursor of a cell that has neighbors with
arbitrary depths, in a $3$-deep, $2$-dimensional binary AMR mesh.} 
\label{fig:HyperTreeBinary2D-SupercursorV2}
\end{figure}
This information can be readily retrieved when using a \emph{Von Neumann
supercursor}, as illustrated in
Figure~\ref{fig:HyperTreeBinary2D-SupercursorV2} (cf.~\cite{harel:17}
for details).
Once again, we start processing the tree from the root level. 
If the cell that we are processing is a non-leaf cell we recursively
process its children.
Once arrived at a leaf cell, each of its 6 faces (as the cell is then
viewed as a rectangular prism) is analyzed: if a face belongs
to an unmasked cell and has no neighbor or the neighbor is masked,
then that face belongs to the surface to be generated. 
In this case, its geometry and topology are computed and appended to
the list of quadrangles defining the output geometry.
Note that the $3$-dimensional case of this filter is identical to that
of the \texttt{Geometry} filter which we implemented, as renderer
information is taken into account in this case.

\subsection{Selection Extraction Filter}
\label{s:extract-selected-filters}
In order to extend the set of filters operating on hypertree
grids to include data analysis capabilities, we started by
extending the selection and extraction methodology of \VTK{} so it can
natively operate on hypertree grids.
We recall that extraction filters are designed to generate subsets
of cells from an input mesh, based on some selection process.
In terms of implementation, the object to be processed is sent to the
first input \emph{port} of the filter, whereas the \emph{selection},
i.e. an instance of a \texttt{vtkSelection} object, is given to a
second port.
This is a first difference with all hypertree grid  filters that we
implemented so far, which operate on a single input, and therefore
only make use of a single input port.
In addition, we wanted to support two different kind of selection modes:
by location (i.e., a geometric criterion) and by identifier (i.e., a
topological criterion).

When the selection mode is by location, the user must provides a list of
coordinates.
The algorithm then recursively explores the tree using a geometric
cursor until it reaches a leaf cell, again using DFS traversal.
In this mode of operation, only non-masked leaf cells can  be
selected, so all others are ignored.
A geometric cursor is also used, so that size and position of
selected cells can be easily retrieved.
The algorithm then checks whether any of these locations are
contained inside each of the traversed cell limits.
In this case, the cell is selected, and  the filter can return two
kinds of outputs: if the \texttt{PreserveTopology} flag is set to
true, then the selection process is made by means of a mask array
attached to the grid.
We recall that such an array associates a value for every cell of the
input hyper tree grid, with value of $1$ for selected cells and a
value of $0$ for non-selected cells. 
On the other hand, when the \texttt{PreserveTopology} flag is not set,
the output result is an unstructured grid containing all selected
cells. 
In the $2$-dimensional case, this output grid is made of
\texttt{VTK\_QUAD} elements, whereas in the $3$-dimensional it only contains
\texttt{VTK\_HEXAHEDRON} cells.

The other mode of operation of the filter is when the
\texttt{vtkSelection} is a list of identifiers (Ids).
In this case, the algorithm is going to search the tree, looking for
these Ids.
Starting from the root cells of the hypertree grid, each constituting
tree is traversed in DFS order.
Given a cell that is not masked by the material mask array, we compare
its global Id with the values contained in the list of identifiers.
If the cell is in the array, it should be selected, otherwise we
keep recursively searching through its children.
We use again a geometric cursor because we need to compute the
coordinates of the points of the selected cells, as with the other
mode of operation.
Here also, the output of the filter is an unstructured grid containing
\texttt{VTK\_QUAD} cells in the $2$-dimensional case, and
\texttt{VTK\_HEXAHEDRON} cells in the $3$-dimensional case.

%% file: 80.tex
\section{Results}
\label{s:results}
We now discuss the main results obtained with our implementation in
\VTK{} of the previously described new hypertree grid filters.
\subsection{Adaptive Surface Filter}
\label{s:adaptive-surface-filter-results}
We begin with the adaptive geometry extraction filter introduced
in~\S\ref{s:adaptive-surface-filter} and implemented 
in~\VTK{}, exploring the $2$ and $3$-dimensional cases as well as
the two possible branch factor values.
\begin{figure}[h!]
\centering
\begin{minipage}[t]{0.48\columnwidth}
\centering
\vspace{0pt}
\includegraphics[width=0.99\columnwidth]{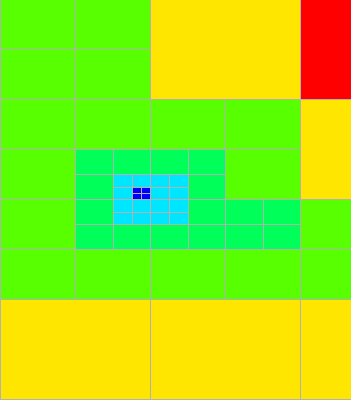}\\
(a)
\end{minipage}
\hfil
\begin{minipage}[t]{0.48\columnwidth}
\centering
\vspace{0pt}
\includegraphics[width=0.99\columnwidth]{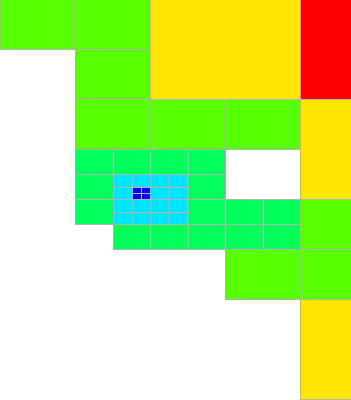}\\
(b)
\end{minipage}
\caption{Visualizations of a part of the data sets produced by the
application of the \texttt{AdaptiveDataSetSurface} filter to a
$2$-dimensional, binary hypertree grid with 6 root cells, without (a)
or with (b) a material mask; only cells that are visible based on
camera and object position are generated.}
\label{fig:HyperTreeGridBinary2DAdaptiveDataSetSurfaceFilter}
\end{figure}

Figure~\ref{fig:HyperTreeGridBinary2DAdaptiveDataSetSurfaceFilter}
displays renderings obtained by applying the
\texttt{AdaptiveDataSetSurface} filter to a 2-dimensional binary hypertree grid,
with a $2\times3$ layout of root cells, to which is attached a single
attribute field filled with the cell depths, either without (a) or
with (b) a non-empty material mask attached to it.
We acknowledge that the adaptive surface filter did indeed
compute a reduced number of polygons, with respect to the non-adaptive
\texttt{Geometry} filter: specifically, where the entire
$2$-dimensional AMR mesh has $75$ leaf cells, the output of
\texttt{AdaptiveDataSetSurface} contains only $67$ quadrangles.
When a material mask is present, the corresponding
numbers are $62$ and $54$, respectively.
In the latter case, the gain is inferior to that of the former, as a
result of the fact that most culling is due to the masking rather than
to the camera position relative to object.

\begin{figure}[ht!]
\centering
\begin{minipage}[t]{0.99\columnwidth}
\includegraphics[width=\columnwidth]{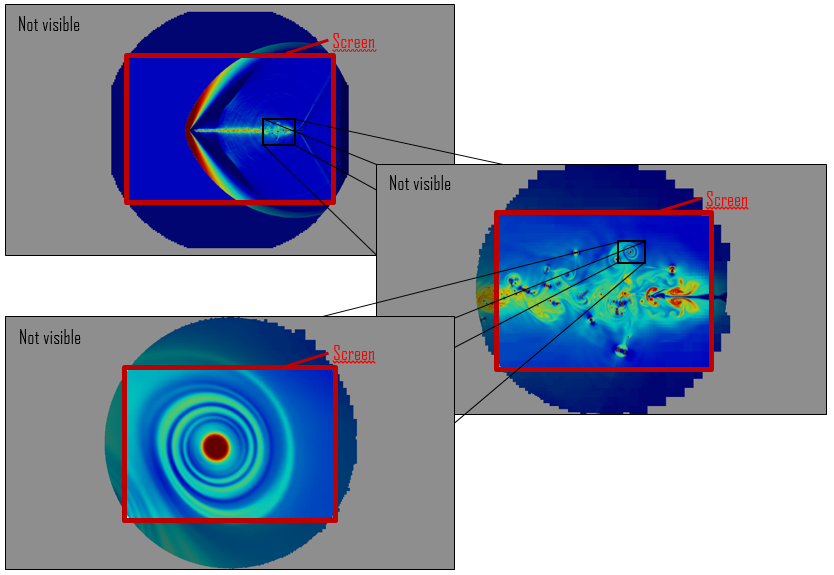}
\end{minipage}
\caption{Successive close-up renderings of a $2$-dimensional AMR
grid obtained with the \texttt{AdaptiveDataSetSurface} filter.}
\label{fig:AMRZoom}
\end{figure}
The elimination of those polygons that are outside the rendering
window or, more precisely, their non-creation, is further illustrated
in Figure~\ref{fig:AMRZoom} with the example of a $2$-dimensional
hypertree grid data set resulting from a computational fluid dynamics
simulation.

\begin{figure}[ht!]
\centering
\begin{minipage}[t]{0.99\columnwidth}
\includegraphics[width=\columnwidth]{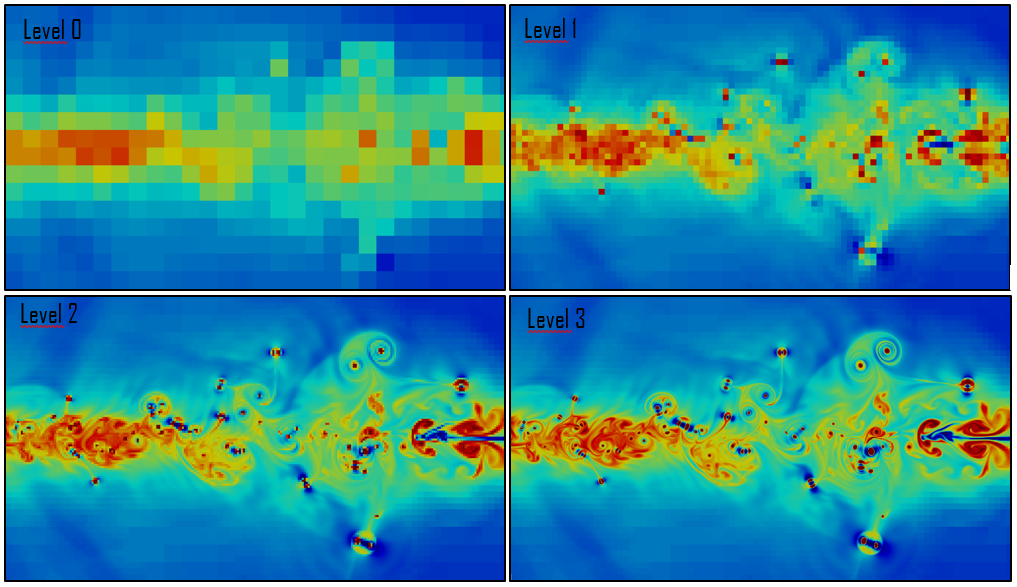}
\end{minipage}
\caption{Renderings of a $2$-dimensional AMR grid obtained with the
\texttt{AdaptiveDataSetSurface} filter, showing the effect of a
decreasing pixelization threshold.} 
\label{fig:levels}
\end{figure}
The sub-pixel culling effect of the adaptive surface filter is
illustrated in Figure~\ref{fig:levels}, again with a $2$-dimensional
hypertree grid, where we can see the effect of adjusting the scale
factor~$s$ in~\eqref{eq:level-max}:
\begin{eqnarray*}
s_1 > s_2 & \Leftrightarrow & - \log{s_1} < - \log{s_2} \\
 & \Leftrightarrow & \delta_{\max}(w,z,s_1,f) <
\delta_{\max}(w,z,s_2,f)\ .
\end{eqnarray*}
In other words, increasing the scale factor results in decreasing the
maximum allowable depth in the DFS traversals of each of the
constituting hypertrees, meaning that more down-sampling will occur.
The same effect obviously occurs when instead of increasing $s$,
either the window size $w$ is decreased, or the view is zoomed out.
\begin{figure}[h!]
\centering
\begin{minipage}[t]{0.48\columnwidth}
\centering
\vspace{0pt}
\includegraphics[width=0.99\columnwidth]{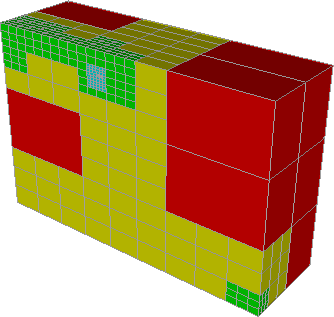}
\end{minipage}
\hfil
\begin{minipage}[t]{0.48\columnwidth}
\centering
\vspace{0pt}
\includegraphics[width=0.99\columnwidth]{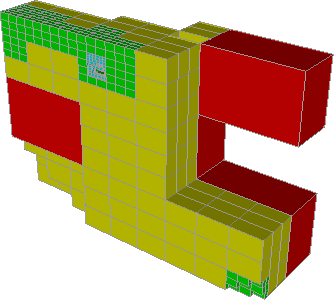}
\end{minipage}
\caption{Renderings of the outside surface of $3$-dimensional, ternary
hypertree grid with 18 root cells, using the
\texttt{AdaptiveDataSetSurface}, without (left) or with (right) a non-empty
material mask.}
\label{fig:HyperTreeGridTernary3DAdaptiveDataSetSurfaceFilter}
\end{figure}

A $3$-dimensional, ternary set of test cases is now used,
with a $3\times3\times2$ layout of roots to further illustrate our point.
We illustrate this in
Figure~\ref{fig:HyperTreeGridTernary3DAdaptiveDataSetSurfaceFilter},
without or with the presence of a non-empty material mask.
We note in particular that in this case, the visualizations obtained
using the novel \texttt{AdaptiveDataSetSurface} filter are indeed
identical to those obtained with the \texttt{Geometry} filter,
cf.~\cite{harel:17} for details; in addition, no culling in the
surface geometry occurred as expected.
\subsection{Selection Extraction Filters}
\begin{figure}[h!]
\centering
\begin{minipage}[t]{0.48\columnwidth}
\centering
\vspace{0pt}
\includegraphics[width=0.99\columnwidth]{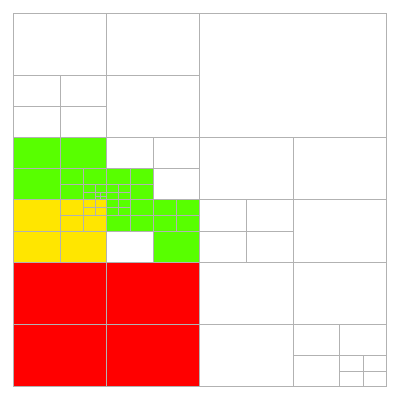}\\
(a)
\includegraphics[width=0.99\columnwidth]{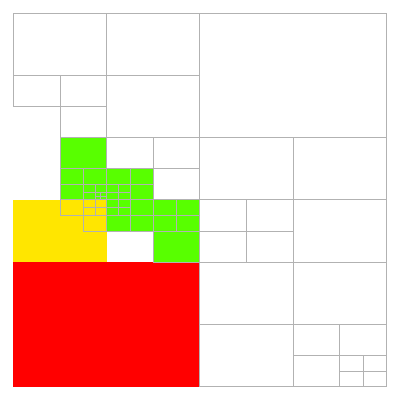}\\
(c)
\end{minipage}
\hfil
\begin{minipage}[t]{0.48\columnwidth}
\centering
\vspace{0pt}
\includegraphics[width=0.99\columnwidth]{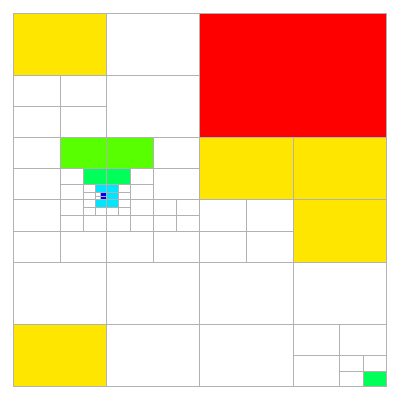}\\
(b)
\includegraphics[width=0.99\columnwidth]{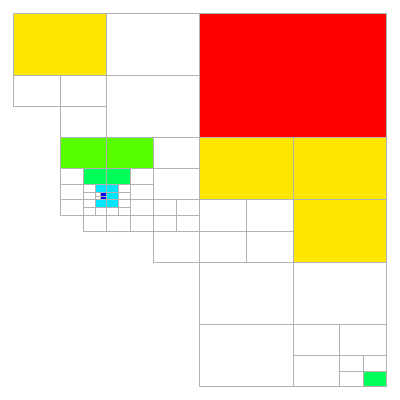}\\
(d)
\end{minipage}
\caption{Visualizations of a part of the unstructured data sets
produced by the application of the \texttt{ExtractSelectedIds} (a\&c) and
\texttt{ExtractSelectedLocations} (b\&d) filters to a
$2$-dimensional, binary hypertree grid with 6 root cells, without
(a\&b) or with (c\&d) a material mask.}
\label{fig:HyperTreeGridBinary2DDataSetSurfaceFilter}
\end{figure}

We continue with the selection extraction filter introduced
in~\S\ref{s:extract-selected-filters} and also implemented in~\VTK{},
exploring the $2$ and $3$-dimensional cases as well as the two
possible branch factor values.
Using the same 2-dimensional binary hypertree grid with a $2\times3$
layout of root cells, as in~\S\ref{s:adaptive-surface-filter-results}
Figure~\ref{fig:HyperTreeGridBinary2DDataSetSurfaceFilter} displays
the renderings obtained with the selection extraction filters applied
to this hypertree grid, either with or without a non-empty
material mask attached to it.

\begin{figure}[h!]
\centering
\begin{minipage}[t]{0.48\columnwidth}
\centering
\vspace{0pt}
\includegraphics[width=0.99\columnwidth]{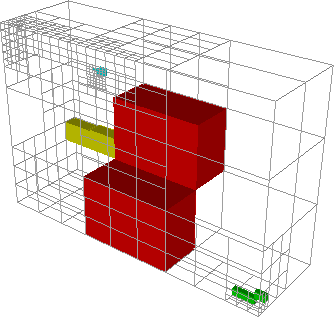}\\
(a)\\[2mm]
\includegraphics[width=0.99\columnwidth]{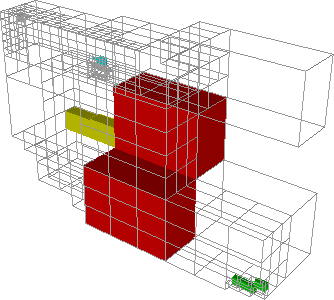}\\
(c)
\end{minipage}
\hfil
\begin{minipage}[t]{0.48\columnwidth}
\centering
\vspace{0pt}
\includegraphics[width=0.99\columnwidth]{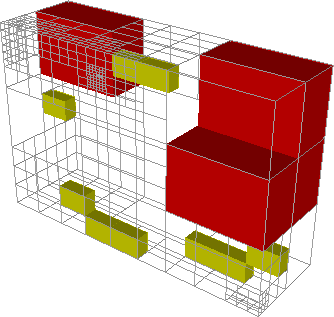}\\
(b)\\[2mm]
\includegraphics[width=0.99\columnwidth]{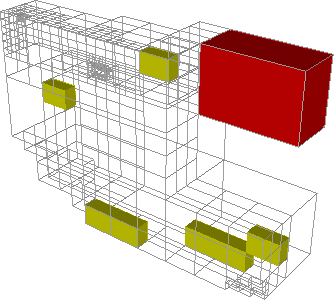}\\
(d)
\end{minipage}
\caption{Visualizations of a part of the unstructured data sets
produced by the application of the \texttt{ExtractSelectedIds} (a\&c) and
\texttt{ExtractSelectedLocations} (b\&d) filters to a
$3$-dimensional, ternary hypertree grid with 18 root cells, without
(a\&b) or with (c\&d) a material mask.}
\label{fig:HyperTreeGridTernary3DExtractSelectedFilter}
\end{figure}

Finally, the same $3$-dimensional, ternary hypertree grid with a
$3\times3\times2$ layout of root cells as
in~\S\ref{s:adaptive-surface-filter-results} is used to further
illustrate our point in
Figure~\ref{fig:HyperTreeGridTernary3DExtractSelectedFilter}.
Furthermore, these baseline comparisons are done with the presence of
a non-empty material mask, as well as without.

The interested reader is invited to draw parallels between
corresponding $2$ and $3$ dimensional sub-figures, and to inspect the
contents of the test harness we implemented for all existing hypertree
grid filters, across different dimensions and branching factors:
8 new individual tests are available and can be either executed
as they are, or modified and experimented with at will.
Of particular interest is the behavior of these
filters in the presence of a material mask, allowing for the
selection of masked out cells.
This can be seen in both $2$ and $3$-dimensional cases and is not a
bug, but a desired feature.

%% file: 90.tex
\section{Conclusion}
\label{s:conclusion}
In this article we provided a description of the new native hypertree
grid filters which we developed, in order to further advance the
vision which we had outlined in~\cite{harel:17}.
One of these two filters implements an efficient version of a commonly
used visualization technique, whereas the two other ones already
belong to the field of data analysis: indeed, these offer the ability
to drill into simulation data sets, and extract from them those parts
that are considered relevant either topologically or geometrically.
These three new filters have been incorporated into the set of native
hypertree grid filters, and we are planning to release them as part of
the open-source \VTK{} library.

For future work, we are considering on one hand to extend the
\texttt{AdaptiveDataSetSurface} filter to support level-of-detailed
rendering for the $3$-dimensional case as well.
Note that we first focused on the $2$-dimensional implementation
first, because this is the case where performance gains are the
highest in relative terms.
This is a result of the fact that, by definition, no culling occurs
when showing the outside surface of planar data set, whereas in
dimension~$3$ only boundary cells can produce surface rectangles.
On the other hand, another selection extraction filter, using
attribute values instead of data set geometry or topology, would be a
useful addition the current data analysis capabilities of the tool set.